\documentclass[aps,pre,12pt,showpacs,draft]{revtex4-1}
\usepackage{graphicx}
\usepackage{amsmath}
\usepackage{default}
\usepackage{default}
\newcommand{\be}{\begin{equation}}
\newcommand{\bea}{\begin{eqnarray}}
\newcommand{\bc}{\begin{center}}            
\newcommand{\ee}{\end{equation}}
\newcommand{\eea}{\end{eqnarray}}
\newcommand{\ec}{\end{center}}
\newcommand{\baa}{\begin{eqnarray*}}
\newcommand{\eaa}{\end{eqnarray*}}
\begin{document}
\title{Near-equilibrium universality and bounds on efficiency \\
 in quasi-static regime with finite source and sink}
\author{Ramandeep S. Johal}
\email{rsjohal@iisermohali.ac.in}
\affiliation{ Department of Physical Sciences, \\ 
Indian Institute of Science Education and Research Mohali,\\  
Sector 81, S.A.S. Nagar, Manauli PO 140306, Punjab, India}
\author{Renuka Rai} 
\email{rren2010@gmail.com}
\affiliation{Department of Applied Sciences,\\ University Institute of
Engineering and Technology \\ Panjab University, Chandigarh-160014, India}
\begin{abstract}
We show the validity of some  results of finite-time thermodynamics,
also within the quasi-static framework of classical thermodynamics.
First, we consider the efficiency at maximum work (EMW)
from finite source and sink modelled as identical
thermodynamic systems. The near-equilibrium
regime is characterized by expanding the internal 
energy upto second order (i.e. upto linear response)  in the difference of
initial entropies of the source and the sink.
It is shown that the efficiency 
is given by a universal expression $2 \eta_C / (4-\eta_C)$,
where $\eta_C$ is the Carnot efficiency. 
Then, different sizes of source and sink are treated, by combining  
different numbers of copies of the same thermodynamic system. The 
efficiency of this process is found to be 
${\boldsymbol\eta}_0 = \eta_C/ (2-\gamma \eta_C)$,
where the parameter $\gamma$ depends only on the relative size of the source
and the sink. This implies that within the linear response theory,
 EMW is bounded as 
${\eta_C}/{2} \le {{\boldsymbol\eta}}_0 \le {\eta_C}/{(2 - \eta_C)}$,
where the upper (lower) bound is obtained with a sink much larger (smaller)
in size than the source. We also remark 
on the behavior of the efficiency beyond linear response.
\end{abstract}
\maketitle
{\bf Introduction:}

Bounds on the  efficiency of idealized thermal processes
have contributed deeply in our understanding of 
laws of nature. For example, Carnot established 
a universal upper bound ($\eta_C$) on the efficiency of heat engines
working between two heat reservoirs, which can 
be achieved by any reversible cycle. Although
Carnot's seminal work dates back to 1824,  
understanding analogous general criteria 
 in finite-time models of thermal 
machines has gained momentum only recently. 
A widely studied quantity in heat engines
is the efficiency at maximum power, $\eta_{\rm mp}$. 
Here the irreversible, finite-rate mechanisms of heat exchange 
have been modelled within various frameworks, like
endoreversible models \cite{Curzon1975, Chen1989}, 
linear irreversible thermodynamics \cite{Broeck2005, Tu2012}, 
stochastic thermodynamics \cite{Schmiedl2008}, low-dissipation 
assumption \cite{Esposito2010} and so on. 
Many of these models \cite{Chen1989, Schmiedl2008, Esposito2010, Tu2012}
obtain the following formula for the efficiency at maximum power:
\be
\eta_{\rm mp} = \frac{\eta_C}{2 - \gamma \eta_C}. 
\label{etamp}
\ee
Here the real parameter $\gamma$ depends on the details
of the particular model and can take values in the range $ 0 \le \gamma \le 1$.  
So the above expression also yields definite 
bounds for $\eta_{\rm mp}$: 
${\eta_C}/{2} \le \eta_{\rm mp} \le {\eta_C}/{(2 - \eta_C)}$ 
(see also \cite{Moreau2011,WangTuEPL, Izumida2012,Broeck2013}).
For instance, under the low-dissipation assumption \cite{Esposito2010},
the upper bound is achieved at maximum
power when the time allocated for the cold contact $\tau_-$
is very small compared to the time for the hot contact $\tau_+$.
The lower bound is achieved for the opposite situation: $\tau_+ \ll \tau_-$.
On the other hand, Ref. \cite{Tu2012} derives these bounds from  
different assumptions based more on the 
principles of linear irreversible thermodynamics. 
Further, in some of these models \cite{Schmiedl2008, Tu2008, Esposito2009},
a universality has been often observed for small values of $\eta_C$,
given by $\eta_{\rm mp} = \eta_C/2 + \eta_{C}^{2}/8$.
While the first order term can be justified for tight-coupling engines within  
linear irreversible thermodynamics \cite{Broeck2005},
 the second-order term is beyond linear response and 
 holds only under further conditions of ``left-right'' 
 symmetry \cite{Esposito2009}. 

Although based on simple models, the comparison of the above bounds 
with efficiencies of real thermal plants looks promising \cite{Esposito2010}.
Still the general conditions, under which these bounds apply, are not clear.
Are these valid only close to equilibrium? The low-dissipation
model is valid close to reversible limit
with long cycle times. On the other hand,
in Ref. \cite{Tu2012} such assumption does not seem
relevant, although one might expect that
validity of linear irreversible thermodynamics
suggests proximity to equilibrium.
In any case, previous approaches involve explicit
time-dependence and an optimization   
over the times of thermal contacts, seems to play 
an important role in the analysis. 

In this letter, we consider a quasi-static framework 
(where, in principle, no time dependence is invoked) 
of work extraction \cite{Thomson, Callen1985, Reifbook, Leff1987b}
from two similar and finite systems acting
respectively as heat source and sink. We show that 
for small temperature differences, the efficiency
at maximum work (EMW) can be shown to be independent of the reservoir
model and depends only on the ratio of initial temperatures.
Further, the bounds on efficiency as mentioned above,  
are also valid for EMW and an 
interpolation between the two bounds is realized 
by taking source and sink of different relative sizes.
Our analysis gives a novel perspective on the validity
of these bounds and the universality of efficiency
in a different, time-independent framework.

First of all, we consider the case of 
two identical thermodynamic systems, each described by 
the same fundamental relation   
$U \equiv U(S, V, N)$. Here $U$, $S$, $V$ and $N$ denote the internal energy,
the thermodynamic entropy, the volume and the number of moles for a system. 
Assume the systems are initially at different temperatures
$T_+$ and $T_-$, where $T_+ > T_-$. Correspondingly, their initial
entropies are denoted as $S_+$ and $S_-$, respectively. 
The simplest case is when $V$ and $N$ are the same
for both systems. We also keep $V$ and $N$ fixed throughout
our discussion.
The total energy is taken as the sum of energies of individual systems. 

Now consider the classic, textbook problem \cite{Callen1985, Reifbook}
in which work is extracted by alternately coupling these systems
with a reversible work source. The total entropy
of the systems is  kept conserved. Thus if $S_1$ and $S_2$ are the 
entropies of the source and the sink at any stage of the process, 
then $S_1 + S_2 = S_+ + S_-$.
Note that no description enters about the working medium 
which is assumed to undergo a cyclic process. 
Classical thermodynamics also tells us that we can extract work
so long as the temperatures of the two systems remain 
different. Due to extraction of heat from the hotter system
and dumping of unutilized heat into the colder system, their 
temperatures gradually approach each other and so the 
process terminates at a common final temperature.
Clearly, as $V$ and $N$ are also same for the two systems,
therefore, in the final state, the 
entropies of both systems are equal, given by $S_c = (S_+ + S_-) /2$. 

Therefore, the maximum extractable work from the systems, due 
to temperature gradient, defined as the difference of initial
and final total energies, is
\be
W_0 = U(S_+,V,N) + U(S_-,V,N) - 2 U(S_c,V,N).
\label{work}
\ee
Note that for an arbitrary reversible process between the 
initial and final states, the notion of change in availability \cite{Gibbs, Callen1985ibid}
encapsulates the useful work. Here, for simplicity, we assume
that no work is performed by/on the source and the sink, 
so that the extracted work is simply given by total change in
internal energies of the systems.

Then the heat absorbed by the work source from the  
hot system is $Q_+ = U(S_+,V,N) - U(S_c,V,N)$. Now we are interested 
in the efficiency
of this process, $\eta_0 = W_0/Q_+$.
We shall prove that if the initial temperature
difference is small, then the efficiency is 
independent of the nature of the source or the sink. 

We make use of an important property
of thermodynamic functions, by which the  
energy is a homogeneous function  of degree one of  
its arguments \cite{Callen1985ibid2}. This implies: $U(\alpha S,\alpha V, \alpha N)
= \alpha U(S,V,N)$,
where $\alpha$ is a scaling factor \cite{Commenthomogeneous}.
Then the efficiency can be written as
\be
\eta_0 = \frac{U(S_+,V,N) + U(S_-,V,N) - U(S_+ + S_-,2V,2N)}
{U(S_+) - \frac{1}{2}U(S_+ + S_-,2V,2N)}.
\label{eta}
\ee
Now we assume that the initial temperature difference $\delta T = T_+ - T_-$ is
small. 
This implies that the entropy difference $\delta S = S_+ - S_-$ 
is small too. Treating entropy as the basic variable, 
we expand internal energy as a Taylor's series upto second-order in $\delta S$: 
\be 
U(S_-,V,N) = U(S_+,V,N) + \left. \frac{\partial U}{\partial S} \right|_{S=S_+}\hspace{-5mm} (-\delta S)
             + \frac{1}{2!} \left. \frac{\partial^2 U}{\partial S^2} 
             \right|_{S=S_+}\hspace{-5mm} (\delta S)^2
             \label{usm}
\ee
and
\bea
U(S_+ + S_-,2V,2N) &= & U(2S_+ - \delta S,2V,2N) \nonumber \\
             &= &  U(2 S_+,2V,2N) + \left. \frac{\partial U}{\partial S} 
             \right|_{S= 2 S_+}\hspace{-5mm} (-\delta S) 
             + \frac{1}{2!} \left. \frac{\partial^2 U}{\partial S^2} 
             \right|_{S= 2 S_+}\hspace{-5mm} (\delta S)^2,
\label{uspm}
             \eea
where all partial derivatives are evaluated at fixed values of $V$ and $N$.
Now notice that 
\be 
T_+ = T_+(S_+,V,N) = \left. \frac{\partial U}{\partial S} \right|_{S= S_+} = \left. 
\frac{\partial U}{\partial S} \right|_{S= 2 S_+}
\ee
which is the intensive property of temperature defined as 
$T = {\partial U}/{\partial S}$.
Secondly, we have
\be
\left. \frac{\partial^2 U}{\partial S^2} \right|_{S= 2 S_+} = 
\left. \frac{\partial T}{\partial S} \right|_{S= 2 S_+} = 
\frac{1}{2}\left. \frac{\partial T}{\partial S} \right|_{S=  S_+}.
\ee
For brevity of the notation, we hide the symbols for $V$ and $N$.
Then using the above conditions, we can write Eqs. (\ref{usm}) and (\ref{uspm}) as 
\be 
U(S_-) = U(S_+) -T_+ \delta S
             + \frac{1}{2} \left. \frac{\partial T}{\partial S} \right|_{S= S_+}
             \hspace{-5mm}(\delta S)^2 
\label{usm2}
             \ee
and
\be 
U(S_+ + S_-) 
             =   U(2 S_+) - T_+ \delta S
             + \frac{1}{4} \left. \frac{\partial T}{\partial S} \right|_{S= S_+} 
             \hspace{-5mm} (\delta S)^2.
\label{u2s}
             \ee
Now, we substitute Eqs. (\ref{usm2}) and (\ref{u2s}) in (\ref{eta}).
Denoting $({\partial T}/{\partial S} )_{S= S_+} \equiv {\cal T}$ and upon
simplifying, we get an expression for EMW as:
\be 
\eta_0 = \frac{ 2 {\cal T} \delta S}{4 T_+ - {\cal T} \delta S}.
\ee
We can identify $ {\cal T} \delta S = \delta T$, whereby 
\be
\eta_0 = \frac{2\delta T}{ T_+ \left (4- \frac{\delta T}{T_+} \right)}. 
\ee
This can be rewritten in terms of Carnot efficiency $\eta_C = \delta T/ T_+$, as
\be
\eta_0 = \frac{2 \eta_C}{4-\eta_C}.
\label{mn1}
\ee
The behavior of EMW, based on the temperature gradient between two finite and 
similar thermodynamic systems, 
is universal in that it depends only on the ratio
of initial temperatures and is independent of the nature of the systems
modelled as heat source and sink. 
This constitutes the first main result of this letter.
To make an analogy, it is interesting to note that 
the above expression for efficiency is also obtained at
maximum power output in stochastic engines \cite{Schmiedl2008}
as well as in exoreversible models \cite{Apertet}.
 
To further see how this efficiency generalizes to different-sized heat source and sink,
we consider $m$ copies of the source-system (hereafter referred to as subsystem)
 in mutual equilibrium at initial temperature $T_+$. 
These subsystems taken together constitute the heat source. Similarly, let $n$ copies of
the subsystem at initial temperature $T_-$ together make up the heat sink. 
Again $S_+$ ($S_-$) is the initial entropy of a single  
subsystem, now comprising the source (sink). Thus $\{mS_+, mV, mN \}$ and 
$\{nS_-,nV,nN\}$ are the values of the corresponding quantities for 
source and sink, respectively. In general, we can take 
$m$ and $n$ to be any real numbers. 

As before, we consider a reversible process in which 
the total entropy is conserved at any stage. This implies that:
$m S_1 + n S_2 = m S_+ + n S_-$.
Finally, the two systems achieve a common temperature.
The entropy of every subsystem in the final state is:
\be
{\cal S}_c =  \frac{m}{m+n} S_+ +  \frac{n}{m+n}  S_-.
\ee
Using the homogeneous property of the energy function $U(S)$, the 
maximum extracted work is given by 
\bea
{\cal W}_0 &=& m U(S_+) + n U(S_-) - (m+n) U(S_c) \nonumber \\
    &=&  m U(S_+) + n U(S_-) - U(m S_+ + n S_-).
    \eea
Heat absorbed from the source by the engine is
\be
{\cal Q}_+ = m U(S_+) - m U(S_c).
\ee
Then the efficiency of this process, ${\boldsymbol\eta}_0 ={\cal W}_0/{\cal Q}_+$, is
\be
{\boldsymbol\eta}_0 =  \frac{m U(S_+) + n U(S_-) - U(m S_+ + n S_-)}{m U(S_+) - 
\left(\frac{m}{m+n}\right)  U(m S_+ + n S_-)}.
\label{etamn}
\ee
Again we assume a small difference in the initial temperatures
of the source and the sink, whereby 
the entropy difference $\delta S=S_+-S_-$ is also
small. Then we have the series expansion of $U(S_-)$ about $S = S_+$, from Eq. (\ref{usm}).
Analogous to Eq. (\ref{uspm}), we have the following series upto second order:
\bea
U(m S_+ + n S_-) &= & U( (m+n) S_+ - n \delta S) \nonumber \\
             &= &  U((m+n) S_+) + \left. \frac{\partial U}{\partial S} 
             \right|_{S= (m+n)S_+} \hspace{-10mm}
             (-n \delta S) 
             + \frac{1}{2!} \left. \frac{\partial^2 U}{\partial S^2} 
             \right|_{S= (m+n) S_+} \hspace{-10mm} (n \delta S)^2.
\eea
The above series can be simplified to
\be 
U(m S_+ + n S_-) 
             =   (m+n) U(S_+) - n T_+ \delta S
             + \frac{n^2}{2(m+n)} {\cal T} (\delta S)^2.            
\label{umn}
             \ee
Substituting Eqs. (\ref{usm}) and (\ref{umn}) into Eq. (\ref{etamn}) and upon 
simplifying, we get
\be
{\boldsymbol\eta}_0 =  \frac{\eta_C}{ 2 -  \left(\frac{n}{m+n}\right) \eta_C}.
\label{etamn2}
\ee
The above expression reduces to the previous result for equal-sized systems, when $m=n$.
Further note the similarity of the above expression with Eq. (\ref{etamp}),
although the two results have very different underlying frameworks. 

Now, consider the following limits. 
To model a sink that is much larger than the heat source,
we can take $n\gg m$. In this case, the expression (\ref{etamn2}) tends to 
the limit 
\be
{\boldsymbol \eta}_0 = \frac{\eta_C}{2 - \eta_C}.
\ee
On the other hand, if the source is much larger compared to the sink
($m\gg n$), then we get the limiting value 
\be
{\boldsymbol \eta}_0 = \frac{\eta_C}{2}.
\ee
Thus with different relative sizes of the source and the sink, 
and for small temperature differences, EMW is bounded as:
\be
\frac{\eta_C}{2} \le {{\boldsymbol\eta}}_0 \le \frac{\eta_C}{2 - \eta_C}.
\ee
The lower (upper) bound is obtained with a sink much smaller (larger) in size
as compared to the source. 
The derivation of the above bounds for efficiency, within a quasi-static
framework of work extraction, constitutes our second major result. 

Finally,  we discuss another aspect of the universality of efficiency.
In scenarios of maximum power output, the efficiency 
is often studied through its series expansion 
in terms of the small parameter $\eta_C$. 
Our expressions for efficiency, as in Eqs. (\ref{mn1}) and (\ref{etamn2}), 
are also given in terms of $\eta_C$. However, it is important
to note that our analysis, 
based on the expansions in $\delta S$ upto the second order, 
is at the level of linear response \cite{Gilmore}. 
In order to get the correct behavior of EMW upto 
the second-order term in $\eta_C$, 
we have to expand the relevant thermodynamic quantities upto third order,
i.e. to go beyond the linear response behavior. 
Hereby, we extend the calculation for the case when  both source and sink
obey the same fundamental relation and the extensive variables are
scaled in the ratio $m:n$.
The detailed calculation is presented in the Appendix. The final expansion
of the efficiency can be written as:
\be
\eta = \frac{\eta_C}{2} + \frac{1}{4(1+n/m)}  \left [ \frac{n}{m} + \frac{(1-n/m)}{3}
\left( 1- \frac{T_+ C_+'}{C_+}\right) \right] \eta_{C}^{2} + O[  \eta_{C}^{3}],
\ee
where $C_+, C_+'$ are the heat capacity at constant volume and its derivative 
w.r.t $T$, both evaluated at $T= T_+$. Thus, we note that the first-order 
term is universal.
The second-order term, in general, depends 
upon the relative sizes, as well as the nature of the  system through $C_+$ and $C_+'$.
Here we mention two special cases in order to evaluate the second-order coefficients.
First, if $m=n$, implying that source and sink are of identical scale, 
then we get the universal coefficient of 1/8, irrespective of the nature
of the system. Secondly, for the simple case of perfect gases, 
the heat capacity is independent of the temperature,
so that $C_+' =0$. Thus for such systems, we get
\be
\eta_{\rm op} = \frac{\eta_C}{2} + \frac{1+2x}{12(1+x)}{\eta_{C}^{2}} + O[\eta_{C}^{3}],
\label{blr}
\ee
where $x=n/m$ and in this case, is also equal to the ratio of heat capacities of 
sink to source. This behavior matches with the behavior 
reproduced through a direct calculation, say by expressiong $U$ directly as
function of $T$, $U(T) = C T$, for a perfect gas. 
Actually, for finite but still large systems,
the correction introduced by $C_+'/C_+$ term is rather
tiny, so that to a good approximation, the second-order term
only depends on the relative size $x$, as in  Eq. (\ref{blr}) above. 

To conclude, the generality of Carnot efficiency
lies in the fact that it is independent
of the nature of the working medium and depends
only on the ratio of reservoir temperatures.
As is well known, a reservoir in Carnot-like or any reversible heat cycle
is characterized only
by its fixed temperature. On the other hand, EMW within  
a finite source/sink setup, is expected to depend,
 in general, on the nature of the (finite) reservoirs
through the function $U(S)$. But as seen above,   
with similar source and sink in a linear response framework, 
EMW is function only of the 
ratio of initial temperatures. For different-sized 
systems modelled as copies of the same 
thermodynamic system, we see an additional dependence
on the relative size of the source and the sink.
The efficiency in this scenario is bounded
from above and below; the specific bounds
are approached when one of the systems becomes very large
in comparison with the other.

More interestingly, the above analysis shows that within linear response theory,
the form of efficiency
at maximum work (quasi-static regime) from finite-sized 
heat source and sink, is similar 
to that found at optimal power output  
in heat engines interacting with (infinite) reservoirs in finite time.
This is despite the fact that the two approaches are based
on very different premises. Thus for instance, the quasi-static framework
involves reversible processes while the finite-time models
involves dissipative processes. 
To the best of our knowledge, these bounds have not been noticed
in literature in the context of a quasi-static work extraction,
although their study within the framework
of finite-time thermodynamics, is an active area of research. We also obtain 
the efficiency upto second order in the (initial) temperature difference, 
which is beyond linear response. We verify that 
irrespective of the fundamental relation, the second-order 
term is universal for systems of identical scale. 
It is hoped that the above analysis 
within the quasi-static regime and near-equilibrium situations, will give 
a fresh perspective in terms of the comparison of the universality   
of efficiency and its bounds at optimal work and power extraction. 

RSJ acknowledges financial support from the Department of
Science and Technology, India, under the research project
No. SR/S2/CMP-0047/2010(G).

\section*{Appendix}
The theory of linear response, as applied to small deviations from 
thermodynamic equilibrium, is used to describe the stability of
the equilibrium state \cite{Gilmore}. In the energy representation, the system energy $U$
 is expanded in terms of the fluctuations in other extensive variables
upto second order. In the present context, we have considered 
expansion of $U$ in terms of entropy $S$, upto the second order,
keeping $V$ and $N$ fixed.
Now, to go beyond linear response,  we expand $W$ and $Q_+$ upto third order in $\delta S$: 
\be
Q_+ = \frac{m n}{m+n} {T_+} {\delta S} -  \frac{m n^2}{2(m+n)^2} {\cal T} {(\delta S)}^2 +  
\frac{m n^3}{6(m+n)^3} {\cal T}'{(\delta S)}^3 + O[{(\delta S)}^4],
\label{qplus}
\ee
\be
W_0 = \frac{m n}{2(m+n)}{\cal T} {(\delta S)}^2 +  \frac{m n(m+2n)}{6(m+n)^2}
{\cal T}' {(\delta S)}^3 + O[{(\delta S)}^4],
\label{w0exp}
\ee
where  ${\cal T} =  \partial T / \partial S |_{S=S_+}$, and
${\cal T}' =  \partial^2 T / \partial S ^2 |_{S=S_+}$. 
The efficiency, upto second order, is written as
\bea
\eta &=& a_1 (-n \delta S) + a_2 n^2 {(\delta S)}^2 + O[{(\delta S)}^3],\\
     & \equiv & \frac{\partial \eta}{ \partial S} \delta S + \frac{1}{2!} \frac{\partial^2 \eta}
     {\partial S^2}
     (\delta S)^2 + O[{(\delta S)}^3],
\label{et2nd}
\eea
where the partial derivative denote that $V$ and $N$ are kept fixed.
To evaluate, say the coefficients $a_1$ and $a_2$, we use expansions
(\ref{qplus}), (\ref{w0exp}) and (\ref{et2nd}) in 
the expression $\eta Q_+ = W_0$. Then we  
compare terms in the same powers of ${\delta S}$ and so
obtain the coefficients, as below.
\bea 
\frac{\partial \eta}{\partial S} & \equiv &  - n a_1  = \frac{{\cal T}}{2T_+}, \label{dns} \\
 \frac{\partial^2 \eta}{\partial S^2} & \equiv & 2 n^2 a_2 =  
 \frac{n}{(m+n)T_+} \left[ \frac{{\cal T}^2}{2T_+} - \frac{(m+2n){\cal T}'}{3n} \right]. 
 \label{dns2}
\label{coeff}
\eea
So formally, we have expressed the EMW, 
in terms of deviations in entropy upto second order, Eq. (\ref{et2nd}).
However, a useful expansion for efficiency is in terms of temperature difference.
For that purpose, we note that
in the energy representation, $T\equiv T(S,V,N)$, 
which is the thermal equation of state.  
Clearly, one can also express efficiency in terms of temperature differences, by appropriately
transforming the independent variable from entropy to temperature. 
Care has to be taken here, because whereas  
the first differential is invariant with respect to such  a 
change of variable, the higher order differentials
are not \cite{Matveev}. 
Thus if we transform the variable from $S$ to $T$, then 
\be
\eta = \frac{\partial \eta}{\partial T} \delta T +
\frac{1}{2!} \left[ 
\frac{\partial^2 \eta}{\partial S^2} \left(\frac{\partial S}{\partial T}\right)^2
+ \frac{\partial \eta}{\partial S} \frac{\partial^2 S}{\partial T^2}
\right] (\delta T)^2 + O[(\delta T)^3],
\label{etios}
\ee
where 
\be 
\frac{\partial \eta}{\partial T} = \frac{\partial \eta}{\partial S} 
           \frac{\partial S}{\partial T} = \frac{\partial \eta}{\partial S} 
           \left( \frac{\partial T}{\partial S}\right)^{-1} =  
           \frac{\partial \eta}{\partial S} {\cal T}^{-1}.
           \ee
Note that all derivatives here are to be evaluated
 at the value $T = T_+$ or $S= S_+$.            
Now, due to relation (\ref{dns}), we get a universal first-order term equal to 
$\eta_C /2$, where $\eta_C = \delta T /T_+$. 
The second-order coefficients can be further evaluated by noting that
\be 
 \frac{\partial S}{\partial T} = \frac{C}{T}.
 \ee
 \be
 {\cal T}' = \frac{\partial^2 T}{\partial S^2} = \frac{\partial}{\partial S} \left( \frac{T}{C}
 \right) =  \frac{\partial T}{\partial S} \frac{\partial}{\partial T} \left( \frac{T}{C} \right) =
 \frac{T}{C} \frac{C - T C' }{C^2},
 \ee
 where $C$ is the heat capacity of the system at temperature $T$ and constant volume $V$, and 
 $C' = {\partial C}/{\partial T}$.
 
 Similarly, 
 \be 
  \frac{\partial^2 S}{\partial T^2} = \frac{\partial}{\partial T} \left( \frac{C}{T} \right)
   = \frac{T C' -C }{T^2}.
  \ee
Using the above derivatives in the second-order expansion of efficiency,
Eq. (\ref{etios}) is finally written as:
\be
\eta = \frac{\eta_C}{2} + \frac{1}{4(1+n/m)}  \left [ \frac{n}{m} + \frac{(1-n/m)}{3}
\left( 1- \frac{T_+ C_+'}{C_+}\right) \right] \eta_{C}^{2} + O[  \eta_{C}^{3}],
\ee
where $C_+, C_+'$ represent the quantities at $T= T_+$.


\begin{references}
%
 \bibitem{Curzon1975}  F.L. Curzon and B. Ahlborn,
  Am. J. Phys. \textbf{43}, 22 (1975).
%
 \bibitem{Chen1989} L. Chen and Z. Yan, 
 J. Chem. Phys. {\bf 90}, 3740 (1989).
 %
\bibitem{Broeck2005}  C. Van den Broeck,
Phys. Rev. Lett. {\bf 95}, 190602 (2005).

\bibitem{Tu2012} Y. Wang and Z.C. Tu, 
Phys. Rev. E {\bf 85}, 011127 (2012).
%
  \bibitem{Schmiedl2008} T. Schmiedl and U. Seifert, 
  Europhys. Lett. {\bf 81}, 20003 (2008).
%
  \bibitem{Esposito2010} M. Esposito, R. Kawai, K. Lindenberg, and 
 C. Van den Broeck, 
 Phys. Rev. Lett. {\bf 105}, 150603 (2010).
 %
\bibitem{Moreau2011} M. Moreau, B. Gaveau and L.S. Schulman, 
arXiv: 1112.1293.
%
\bibitem{WangTuEPL} Y. Wang and Z.C. Tu, 
Europhys. Lett. {\bf 98},  40001 (2012). 
%
\bibitem{Izumida2012} Y. Izumida and K. Okuda, 
Europhys. Lett. {\bf 97}, 10004 (2012). 
 %
 \bibitem{Broeck2013} C. Van den Broeck, 
 EPL {\bf 101}, 10006 (2013). 
 %
 \bibitem{Tu2008} Z.C. Tu, 
 J. Phys. A: Math. Theor. {\bf 41}, 312003 (2008).
 %
 \bibitem{Esposito2009} M. Esposito, K. Lindenberg, and 
 C. Van den Broeck, 
 Phys. Rev. Lett. {\bf 102}, 130602 (2009).
 %
 \bibitem{Thomson}  W. Thomson, 
 Philos. Mag. {\bf 5}, 102 (1853). 
 %
 \bibitem{Callen1985} H.B. Callen, {\it Thermodynamics and an Introduction 
to Thermostatistics}, Second edition, John Wiley \& Sons (1985), Chap. 4. 
  %
\bibitem{Reifbook} F. Reif, Fundamentals of Statistical and Thermal Physics,
Problem 5.23, (McGraw-Hill 1981).
%
 \bibitem{Leff1987b}  H.S. Leff, 
 Am. J. Phys. {\bf 55}, 701 (1987).
 %
 \bibitem{Gibbs} J.W. Gibbs, {\it Collected Works, Vol. 1:  Thermodynamics}
 MIT Press, Cambridge, Mass., (1970).
 %
 \bibitem{Callen1985ibid} {\it Ibid.} \cite{Callen1985}, Problem 4.5-20.
 
 \bibitem{Callen1985ibid2} {\it Ibid.} \cite{Callen1985}, Chapter 3.
 \bibitem{Commenthomogeneous} Here, 
 the homogeneous property is implied with respect to scaling of 
 all extensive parameters:  
 $U(\alpha S, \alpha V, \alpha N) = \alpha U(S,V,N)$.  For instance, 
 $2 U(S_c) \equiv 2 U(S_c, V, N) = U(S_+ + S_-, 2V, 2N)$.
 %
\bibitem{Apertet} Y. Apertet, H. Ouerdane, C. Goupil and Ph. Lecoeur,
Phys. Rev. E {\bf 85}, 031116 (2012). 
%
\bibitem{Gilmore} R. Gilmore, {\it The Structure of Thermodynamics}, Lecture Notes,
Drexel University (2008).
%
\bibitem{Matveev}O.V. Manturov and N.M. Matveev, {\it A Course of
Higher Mathematics}, Mir Publishers, Moscow (1989) p. 214.





 %
\end{references}
\end{document}